\documentclass[
  aps,prl,
  10pt,
  superscriptaddress,
  twocolumn, floatfix]{revtex4-1}

\usepackage{natbib}

\usepackage[utf8]{inputenc}
\usepackage[T1]{fontenc}
\usepackage{lmodern}
\usepackage{xcolor}

\definecolor{linkblue}{RGB}{31,119,180}
\usepackage[
  unicode,
  colorlinks,
  citecolor=blue,
  linkcolor=linkblue,
  urlcolor=linkblue,
  bookmarks=true,
  bookmarksopen=true,
  bookmarksopenlevel=3,
  bookmarksnumbered=true]{hyperref}
\usepackage{siunitx}
\usepackage{graphicx}
\usepackage{amsmath}
\usepackage{amssymb}
\usepackage{fontawesome}

\DeclareMathOperator{\sinc}{sinc}

\setcitestyle{super,comma}
\begin{document}

\title{Optical sensitivities of current gravitational wave observatories \\at higher kHz, MHz and GHz frequencies}
\author{R. Schnabel*}
\author{M. Korobko}

\affiliation{Institut f\"ur Quantenphysik \& Zentrum f\"ur Optische Quantentechnologien, Universit\"at Hamburg,\\%
Luruper Chaussee 149, 22761 Hamburg, Germany}

\date{\today}

\begin{abstract}
GEO\,600, Kagra, LIGO, and Virgo were built to observe gravitational waves at frequencies in the audio band, where the highest event rates combined with the largest signal to noise ratios had been predicted. Currently, hypothetical sources of cosmological origin that could have produced signals at higher frequencies are under discussion. What is not widely known is that current interferometric GW observatories have a frequency comb of high optical sensitivity that encompasses these high frequencies. Here we calculate the high-frequency noise spectral densities of operating GW observatories under the justified assumption that photon shot noise is the dominant noise source. We explain the underlying physics of why high sensitivity is achieved for all integer multiples of the free spectral ranges of the observatory's resonators when an interferometer arm is not orientated perpendicular to the propagation direction of the GW. Proposals for new concepts of high-frequency GW detectors must be compared with the high-frequency sensitivities presented here.
\end{abstract}

\maketitle

\section{Introduction}
At the time when Rainer Weiss analysed the concept of earthbound laser interferometric gravitational wave (GW) detection in terms of signal strength and noise more than 50 years ago \cite{Weiss1972}, astrophysical sources of signals in the audio band were known \cite{Forward1967,Ostriker1969}. The probability of being able to measure GW signals on Earth in this band increased further over the following ten years \cite{Rees1973,Epstein1975,Thorne1980,Thorne1987,Schutz1989}.
At the turn of the millennium, a total of six Michelson-type laser-interferometric GW detectors -- GEO\,600~\cite{Willke2002},  LIGO\;(3)~\cite{Abramovici1992}, Tama \cite{Uchiyama1998} and Virgo \cite{Bradaschia1990} -- were under construction, targeting the audio-band. 
On September 14th, 2015, Advanced LIGO observed the first GW, which had frequency components up to about 300\,Hz, emitted by the merger of two black holes at a distance of about 1.3 billion light years \cite{GW150914}. By 2020, up to 90 signals from compact binary mergers were detected by LIGO and Virgo \cite{GWo3b2023}. 

Also {\it below} the audio band, a large number of sources are expected to emit signals of measurable amplitude. Avoiding the strong terrestrial noise in this frequency range, LISA is a space-based GW observatory that targets the range from 0.1\,mHz to 0.1\,Hz \cite{Danzmann1995}. It is due to be launched in the 2030s. Pulsar timing arrays (PTAs) are used to measure GW in the nHz range \cite{Moore2015}. In 2023, several PTA collaborations found evidence for an incoherent background of gravitational waves produced by the collisions and mergers of supermassive black holes, see for instance \cite{Agazie2023}.
\begin{figure}
  \includegraphics[width=0.9\linewidth]{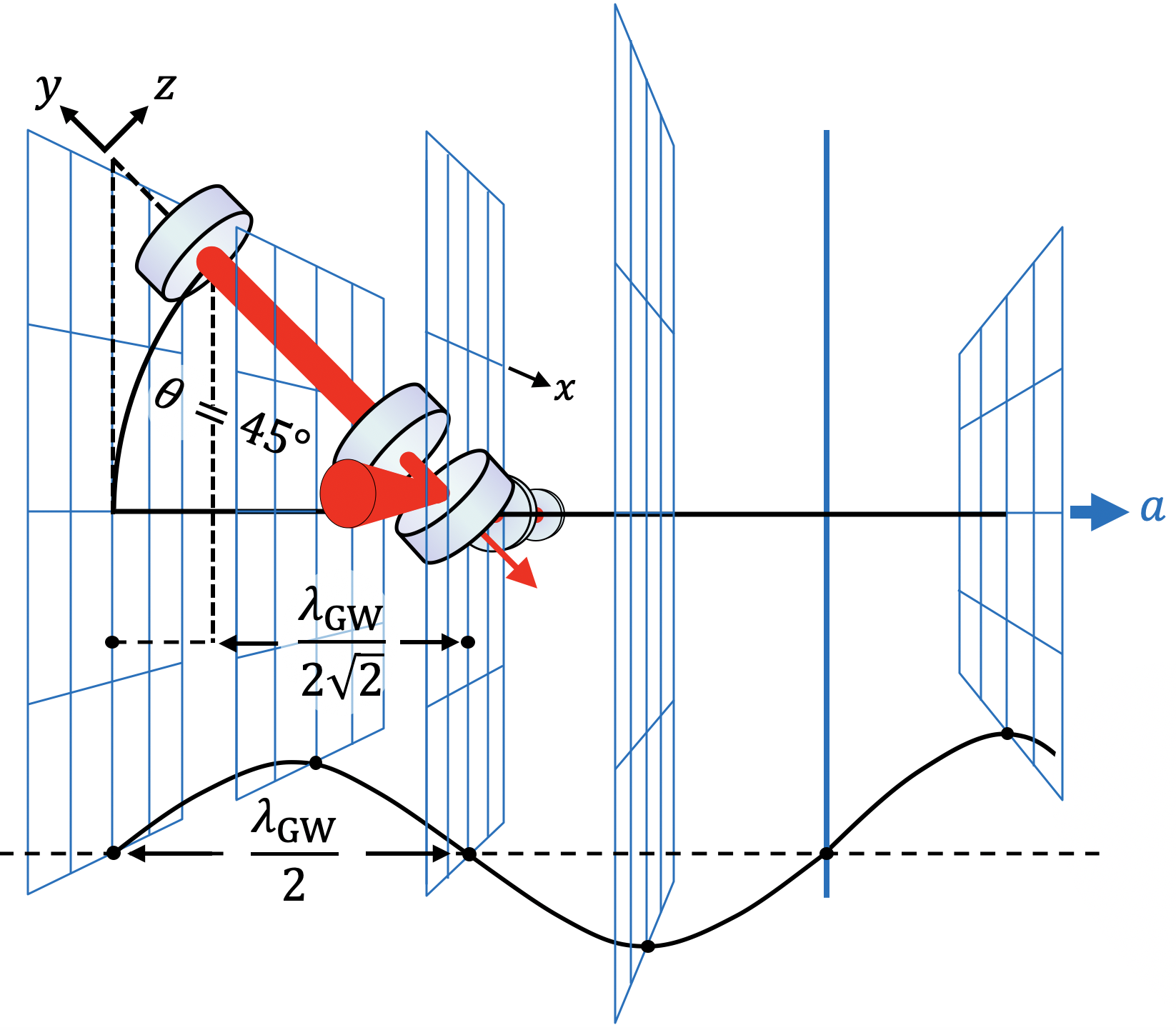}\vspace{-4mm}
  \caption{Illustration of an example alignment for efficiently observing a high-frequency GW propagating along $a$ and having a wavelength that corresponds to the round trip length of the arm resonators of an interferometric GW observatory ($\lambda_{\rm GW} \!=\! 2L$). Here, the arm resonator along the $x$-direction does not sense any signal. However, the arm resonator tilted by $\theta = 45^\circ$ does efficiently pick up the GW, because the GW induced change of the light's propagation length is halved while the geometric length change scales down by just $\sqrt{2}$. The different scaling prevents the complete cancellation of the optical phase signal, which is then resonantly amplified by the resonator.
  }\vspace{-2mm}
  \label{fig:1}
\end{figure}

GWs at frequencies {\it above} the audio band would be rather exotic. There are no known astrophysical sources from star formation or evolution that could emit such high frequencies with measurable amplitudes. However, cosmological sources from the early universe cannot be ruled out. The early universe is not well understood, and inflation \cite{Grishchuk1975}, first-order phase transitions \cite{Caprini2016}, topological defects \cite{Blanco-Pillado2017}, and other effects could have generated gravitational waves that still have frequencies in the MHz or even GHz range today despite redshift. A review can be found in Ref.\,\cite{Caprini2018}.
New detectors have been proposed to measure gravitational waves above 10\,kHz. A recent and highly cited review is given by Ref.\,\cite{Aggarwal2021}. However, it was largely overlooked that today's gravitational wave detectors can measure high-frequency gravitational waves not only via the non-linear memory effect \cite{Christodoulou1991,Favata2010},
but also have a considerable optical sensitivity above 10\,kHz as well as in the MHz and GHz range. 
As early as 2002, it was found that first-generation GW detectors have relatively high sensitivity at frequencies corresponding to integer multiples of the free spectral range (FSR) of the arm resonators  \cite{Rakhmanov2002,Rakhmanov2006TLIGO}. In the case of the LIGO arm resonators, the FSR is 37.5\,kHz. A readout channel for this frequency was developed by down-converting the signal to match the existing LIGO data acquisition system \cite{Markowitz2003TLIGO}. 
The LIGO antenna pattern was analyzed on gravitational-wave frequencies corresponding to multiples of the cavity free spectral range \cite{Elliott2005}.
 
Here, we present the optical strain sensitivities up to $10^{11}$\,Hz of the Advanced LIGO observatory, representing also Virgo and KAGRA  \cite{Akutsu2019}, as well as GEO\,600. We additionally consider a 100-m and a 1-m laser interfero\-meter with high-finesse arm resonators, but no other resonant enhancements such as power or signal recycling. 
At about 1\,MHz, current LIGO achieves a one-sided strain-normalised shot noise spectral density of the order of $10^{-22}/\sqrt{\rm Hz}$. Generally, such high sensitivities at high frequencies are achieved if observatory arm resonators are tilted in the direction of GW propagation and if the GW frequency matches the differential frequency of two longitudinal modes of the optical resonator. The latter condition corresponds to the situation when the arm resonator roundtrip length equals an integer multiple of the GW wavelength. An example is illustrated in Fig.\,\ref{fig:1}.

\section{Metric of a weak gravitational wave}
In all metrics that solve Einstein's field equations, the 4-dimensional events along a propagating laser beam have zero distances. 
The metric of a weak gravitational wave that is $(+)$polarised and propagates along the z-direction is therefore described in the transverse-traceless (TT) gauge by
\begin{equation} 
	ds^2 \!=\! -c^2 dt^2 + (1\!+\!h_+) dx^2 + (1\!-\!h_+) dy^2 + dz^2\!=0 \, ,
\label{eq:1}
\end{equation}
where $|h_+| \!\ll\! 1$ is the amplitude of the polarised GW as described above and $c$ is the speed of light.
Eq.\,(\ref{eq:1}) makes it possible to easily determine the change in the propagation time of a laser beam along the $y$-direction (or $x$-direction)\,\cite{rakhmanov2005response,Rakhmanov2008,Rakhmanov2009}.
The above equation simplifies to
\begin{equation}
  c^2 dt^2 = (1 - h_+(t,y))dy^2 \, .
\end{equation}
The time span $\tau (y)$ that the light needs to propagate from $y_0$ to $y$ is for $|h_+| \!\ll\! 1$ then given by
\begin{align}
	\tau(y) &= t_0 + \frac{1}{c}\int\limits_{y_0}^{y} \sqrt{1-h_+ (\tau_0(y'))} \;dy'  \nonumber \\ 
		   &\approx t_0 + \frac{y-y_0}{c} - \frac{1}{2c}\int\limits_{y_0}^{y} h_+\!\left(\tau_0(y')\right) dy' \, ,
\end{align}
where $\tau_0(y') = t+(y'-y_0)/c$ is the unperturbed time span for light starting at time $t$. 
For a monochromatic GW with amplitude $h_+(t) \!=\! h_+ \!\cos(2\pi f t)$ and a laser beam that is retro-reflected by a mirror at distance $L$, 
the change in the round trip time at $y_0=0$ is given by
\begin{multline}\label{eq:4}
	\Delta \tau_{2L} = -\frac{h_{+}}{2c}\int\limits_0^{L} \cos\!\left[2\pi f\! \left(t+\frac{y'}{c}\right)\right] dy'  \\  \left.
		+ \frac{h_{+}}{2c}\int\limits_L^0 \cos\!\left[2\pi f\! \left(t+\frac{y'-L}{c}\right)\right] dy' \, .  \right.
\end{multline}
The solution of the integral provides the known amplitude for the change in the round-trip time in an arm resonator aligned along $y$ in case of a $(+)$polarized GW
\begin{equation}
	\Delta \tau_{2L} = -\frac{Lh_+}{2c} \sinc\! \left(\frac{\pi f}{f_{\rm FSR}}\right) \, ,
\label{eq:5}
\end{equation}
where $f_{\rm FSR} = c/(2L)$ is the free spectral range of the arm resonator. For an arm resonator in the $x$ direction, Eq.\,(\ref{eq:5}) has the opposite sign. 
For low GW frequencies, i.e.~$f \!\ll\! f_{\rm FSR}$ (audio-band frequencies for km-scale arm resonators) one gets the well-known relation 
\begin{equation}
	\Delta \tau_{2L} \approx -\frac{Lh_+}{2c}   \;\;\;\Rightarrow\;\;\;    \Delta L \approx -\frac{Lh_+}{2} \, ,
\label{eq:6}
\end{equation}
where $\Delta L$ is the amplitude of the effective arm length change.\\

Relevant for this work is Eq.\,(\ref{eq:5}). It states that the time delay due to GW is zero, if the GW frequency is $f = n \!\cdot\! f_{\rm FSR}$, with $n$ a natural number. Often overlooked, however, is the limited range of validity of Eq.\,(\ref{eq:5}), and it is therefore wrongly concluded that laser interferometric GW observatories are generally not sensitive to GWs at these ``FSR-frequencies''.
In fact, the current GWOs are only insensitive to these high-frequency gravitational waves if they come from the zenith (or nadir). For all other alignments, interferometric GW observatories have significant response precisely at all GW frequencies that correspond to an integer multiple of $f_{\rm FSR}$.
The rather high response at these particular frequencies comes from the fact that optical resonators resonantly enhance all signal frequencies $f = n \!\cdot\! f_{\rm FSR}$ since these frequencies correspond to the frequency separation of neighbouring longitudinal resonator modes.

As an example, we calculate the time delay for a resonator round trip when the resonator is inclined at $\theta = 45^\circ$ against the propagation direction of the GW, as sketched in Fig.\,\ref{fig:1}. The coordinate transformation between the $(+)$polarisation and the $(x, y)$-oriented arms leads to a halving of the GW amplitude contribution. Additionally, the trajectory of light changes over the duration of the gravitational wave, i.e.~we replace in Eq.\,(\ref{eq:1}) $h_+(t)$ by $h_+\!\left(t+y/\sqrt{x c^2}\right)/2$ for the light propagating along the $y$ axis.
If we carry out the calculation analogous to the one above, we arrive at a relatively simple integral to solve. 
Here, we limit ourselves to specific GW frequencies. 
For $f\!\ll\!f_{\rm FSR}$ we obtain 
\begin{equation}
	\Delta \tau_{2L} \approx -\frac{Lh_+}{4c}   \;\;\;\Rightarrow\;\;\;    \Delta L \approx -\frac{Lh_+}{4} \, ,
\label{eq:7}
\end{equation}
i.e.~half of the signal for the optimal alignment.\\ 
For $f = f_{\rm FSR}$, the integral reduces to
\begin{equation}
  \!\!\Delta \tau_{2L}= \frac{-1}{4c}\int\limits_0^L \!h_{+}\cos\!\frac{\pi y'}{L} \cos\!\frac{\pi y'}{\sqrt{2}L} \,dy' \approx -\frac{0.18 L}{c} h_+ \,,
\label{eq:8}
\end{equation} 
which is only slightly worse than the response at low frequencies according to Eq.\,(\ref{eq:7}).
Similarly, the detector has significant response at all cavity FSRs, see next section.

Current GW observatories have optical responses to gravitational waves according to Eqs.\,(\ref{eq:5})-(\ref{eq:8}).
Since our calculation uses the TT-gauge, the light's red-shift (when propagating through expanding spacetime) and blue-shift (when propagating through shrinking spacetime) are already included \cite{rakhmanov2005response}.
We note that we have set the phase of the GW to zero. In the more general case, the above equations would get slightly more complex, as we show below.

\section{Frequency combs of high sensitivity of current GW observatories}

An optical resonator shows a longitudinal resonance if its round trip length $2L$ equals an integer multiple of the wavelength of the light coupled to it ($2L = n \cdot \lambda$). The frequency spacing of two neighboring resonances is called the free spectral range ($f_{\rm FSR} = c/2L$). Phase modulations of carrier light that meets one resonance condition are optically enhanced at all frequencies $f$ that correspond to $n \cdot f_{\rm FSR}$, where $n$ is again a natural number.  
In the case of the 4\,km LIGO detectors, the comb spacing is 37.5\,kHz. The 3\,km Virgo and KAGRA detectors have a comb spacing of 50\,kHz. The 1.2\,km long, folded-arm signal recycling cavity of GEO\,600 results in integer multiples of 125\,kHz. In all cases, the linewidths of the resonances are of the order of a kHz. The proposed 10\,km Einstein Telescope and the 40\,km Cosmic Explorer have comb spacings of 15\,kHz and 3.75\,kHz \cite{Essick2017}.

\begin{figure}[t]
  \includegraphics[width=0.98\linewidth]{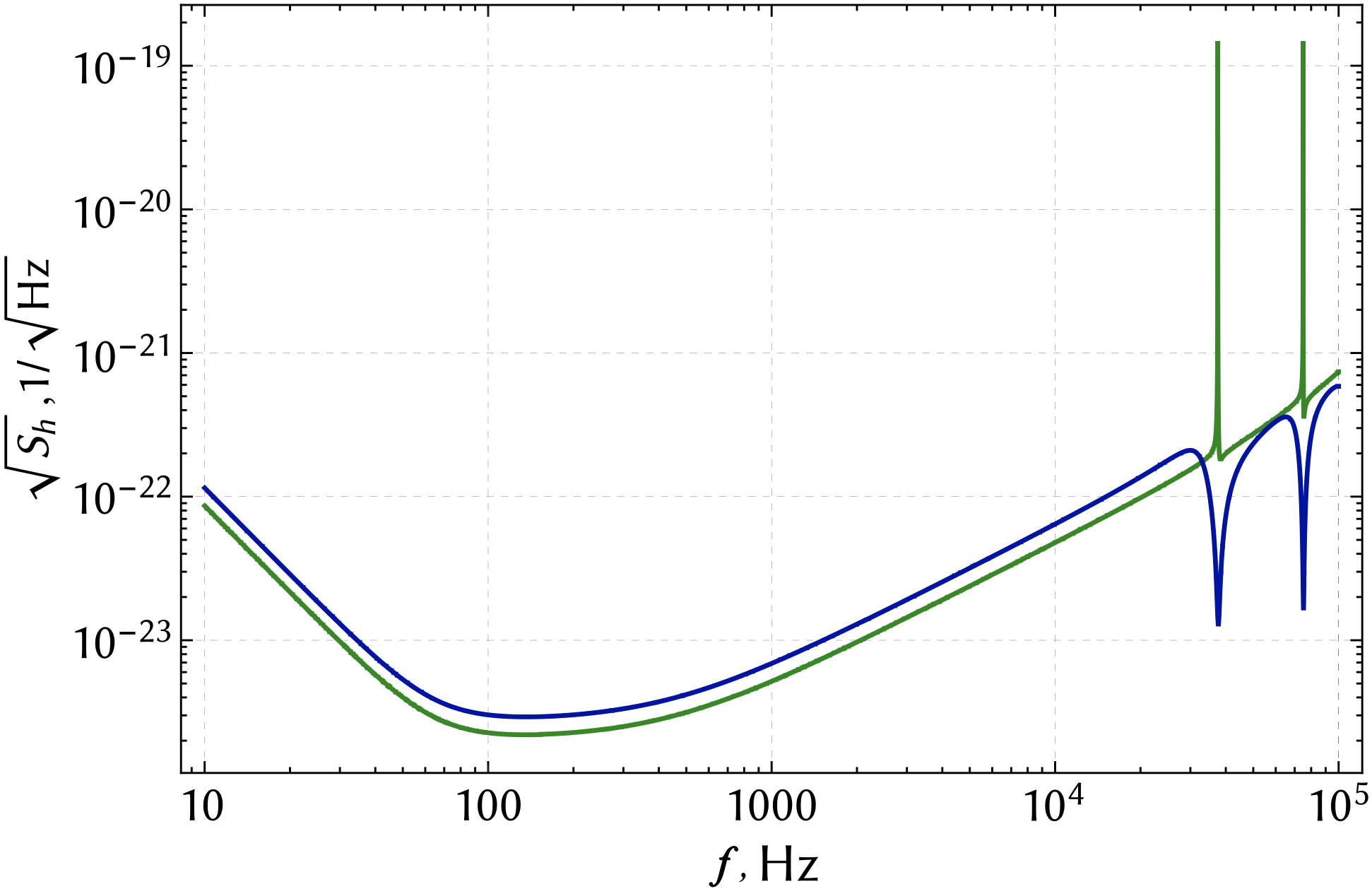}\vspace{-2mm}
   \caption{Strain-normalized quantum noise amplitude spectral densities of Advanced LIGO according to its design values (without quantum noise squeezing \cite{Schnabel2010,Schnabel2017}) for two different sky locations of the source, averaged over $(+)$ and $(\times)$polarisations of the GW. Lowest noise at audio band frequencies is achieved for the zenith sky location (green), which the inverse of Eq.\,(\ref{eq:5}) refers to. The second curve (blue) refers to a sky location that minimizes the noise to signal ratio at LIGO's first ``FSR frequency'' of 37.5\,kHz.
   }\label{fig:2}
 \end{figure}
 
Spectral sensitivities of GW observatories are best described by one-sided (positive frequencies only) amplitude noise spectral densities (ASD) normalized to the signal strength at the respective frequency. The relation between the ASD normalized to strain $h$ and normalized to the round trip phase difference of the laser beams $\varphi$ reads
\begin{align}
  \sqrt{S_{h}(f)} = \frac{c}{2L\omega} \sqrt{S_{\varphi}(f)} \; ,
\label{eq:9}
\end{align}
where $\omega$ is the angular frequency of the laser light. If the noise in $S_\varphi (f)$ is dominated by photon shot noise \cite{Schnabel2017}, which is a well-justified approximation for signal frequencies above a few kilohertz, the noise spectrum alone is ``white'', i.e.~independent of the frequency. The phase signal on the laser light, however, depends on the frequency and additionally on the GW's polarisation with respect to the orientation of the observatory, on its alignment with respect to the GW's direction of propagation, and on length and linewidth of the arm resonators and further enhancement resonators coupled to it.
We show how the phase signal for an observatory with two equally long arms under 90$^\circ$ can be calculated for arbitrary alignments in the next section.

Fig.\,\ref{fig:2} presents quantum noise amplitude spectral densities of Advanced LIGO for two different sky locations of the GW sources for mixed GW polarisation. The zenith sky location provides the lowest noise for audio-band frequencies (green). The second sky location is optimized for lowest noise at LIGO's FSR frequency of 37.5\,kHz (blue). LIGO's quantum noise limited amplitude sensitivity at this frequency is just a factor of about six lower than that at the optimal frequency between 100 and 200\,Hz. 
\begin{figure}[t]
  \includegraphics[width=0.98\linewidth]{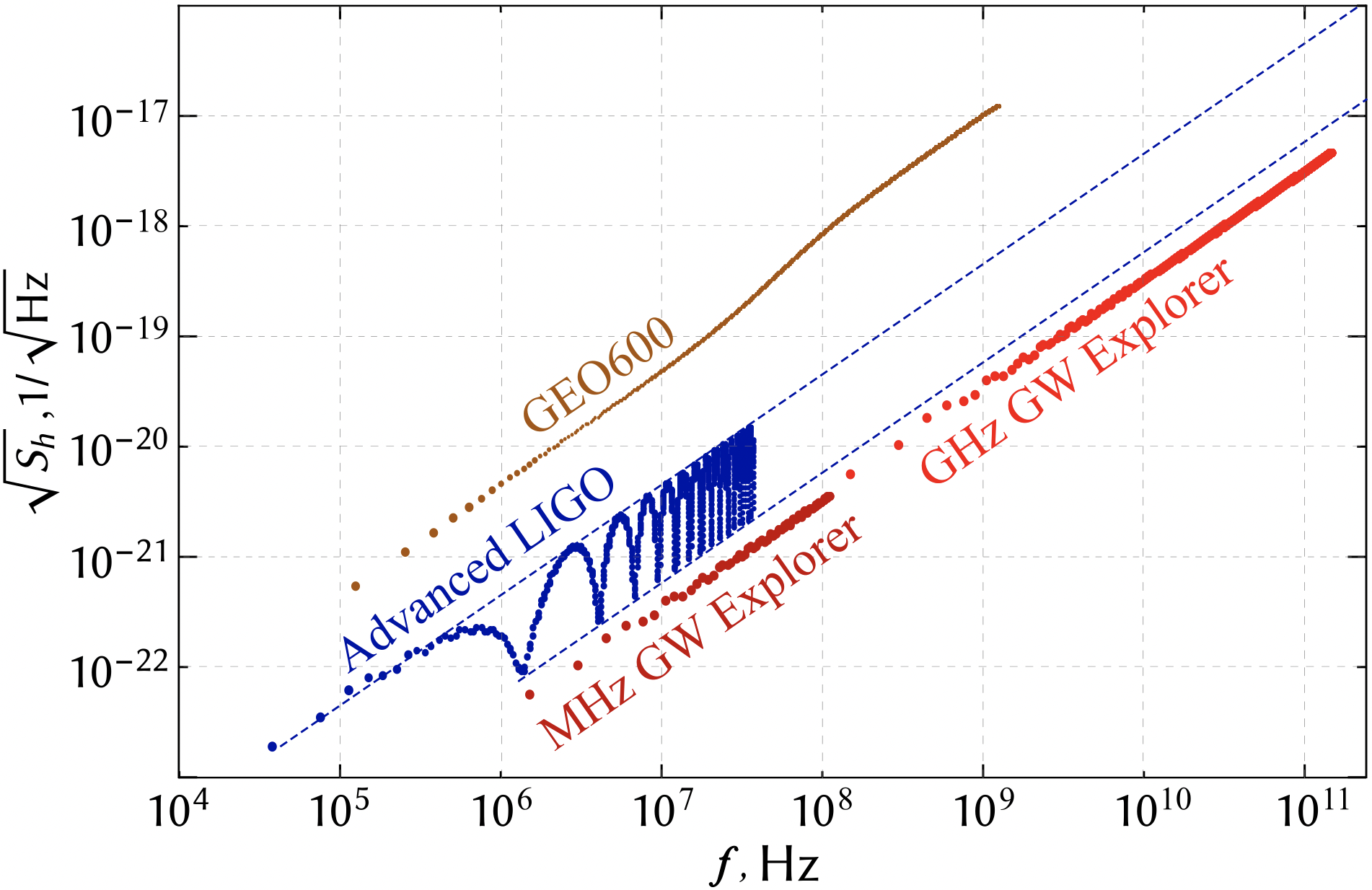}\vspace{-2mm}
   \caption{Polarisation-averaged and sky-averaged amplitude shot-noise spectral densities at frequencies $n\!\cdot\!f_{\rm FSR}$. 
GEO\,600 for up to $n=10^4$: folded 1200-m arms with 1.5\,kW light power in each arm, power recycling resonator, no arm resonators, and signal recycling resonator and 6\,dB quantum noise squeezing \cite{LSC2011,Lough2021} up to about 50\,MHz \cite{Vahlbruch2010}. Advanced LIGO with design values ($n\!\le\!10^3$): power-recycled 4-km arm resonators with 750\,kW each, signal extraction resonator (no quantum noise squeezing) \cite{Aasi2015a}. The noise to signal ratio further drops with a frequency spacing of about 3\,MHz. This is due to alternating change from signal extraction to signal recycling. Dashed lines are the extrapolated envelopes. 
Also shown is our result for two potential small-scale laser interferometers which we name ``MHz-GW explorer'' ($n\!\le\!63$) and ``GHz-GW explorer'' ($n\!\le\!10^3$): 100-m and 1-m arm resonators, respectively, with 10\,MW optical power at 1550\,nm, input mirror reflectivity of 99.995\%, 10\,ppm round trip loss (here no squeezing).  
   }
   \label{fig:3}
 \end{figure}

Fig.\,\ref{fig:3} presents polarisation-averaged and sky-averaged amplitude shot-noise spectral densities at the ``FSR frequencies'' of two GW detectors in operation and two conceivable detectors with less complexity and shorter arms.  
In all traces, the noise to signal ratio increases proportional to the GW frequency. This is a general property of laser interferometers with resonator round trip time larger than the gravitational wave period, because the effective propagation time over which the effect of the GW is accumulated is inversely proportional to the GW frequency. This property is described by the sinc-function in Eq.\,(\ref{eq:5}).
Comparing the interferometer sensitivities shows that for frequencies above the largest FSR (here 150\,MHz), arm length is not an issue. 
\begin{figure}[h!!!!!!!!!!!!!!!!!!!!!!!!!]
  \begin{minipage}{.5\textwidth}
    \includegraphics[width=0.75\textwidth]{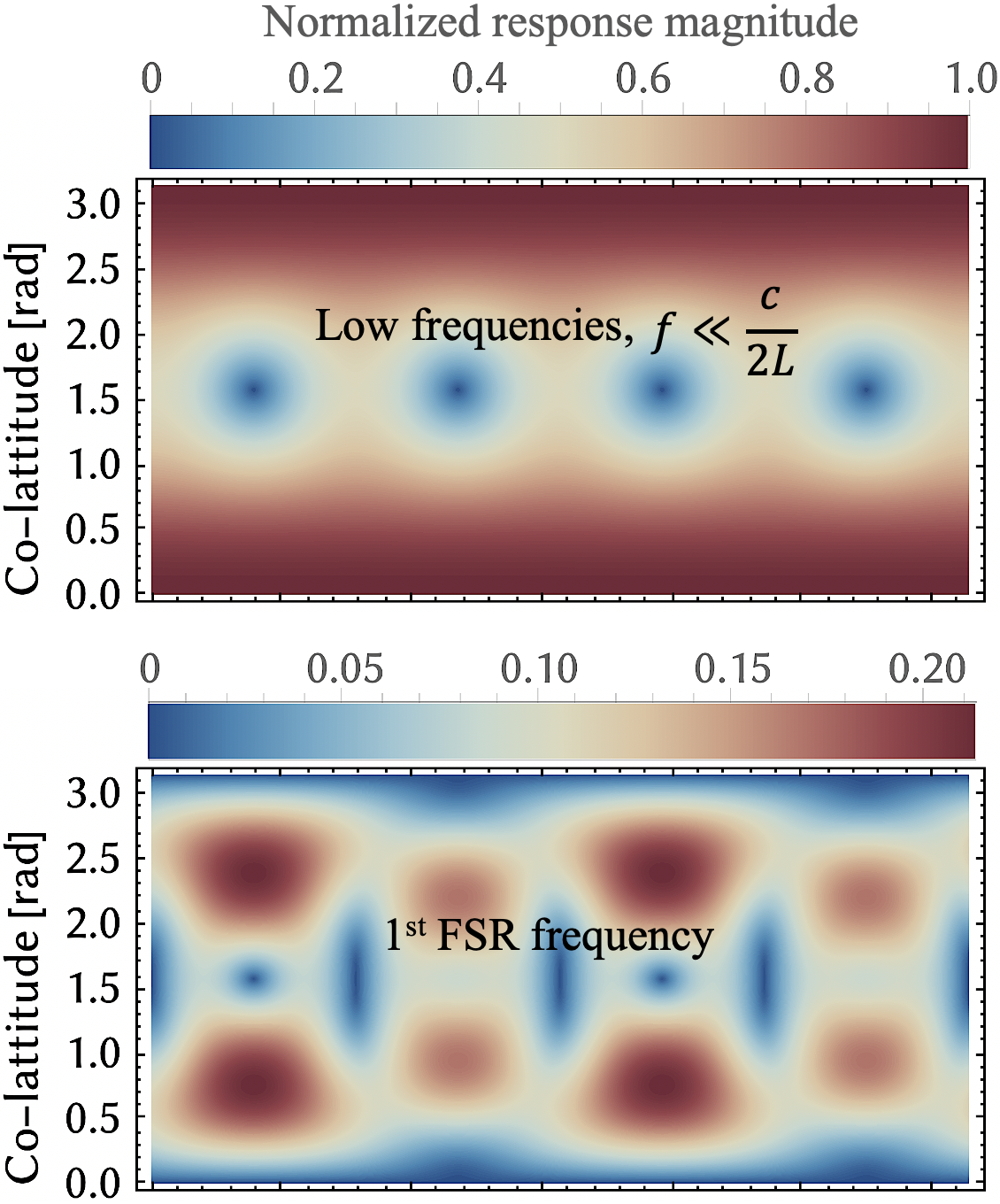}
  \end{minipage}
  \begin{minipage}{.5\textwidth} \vspace{1.5mm}
    \includegraphics[width=0.755\textwidth]{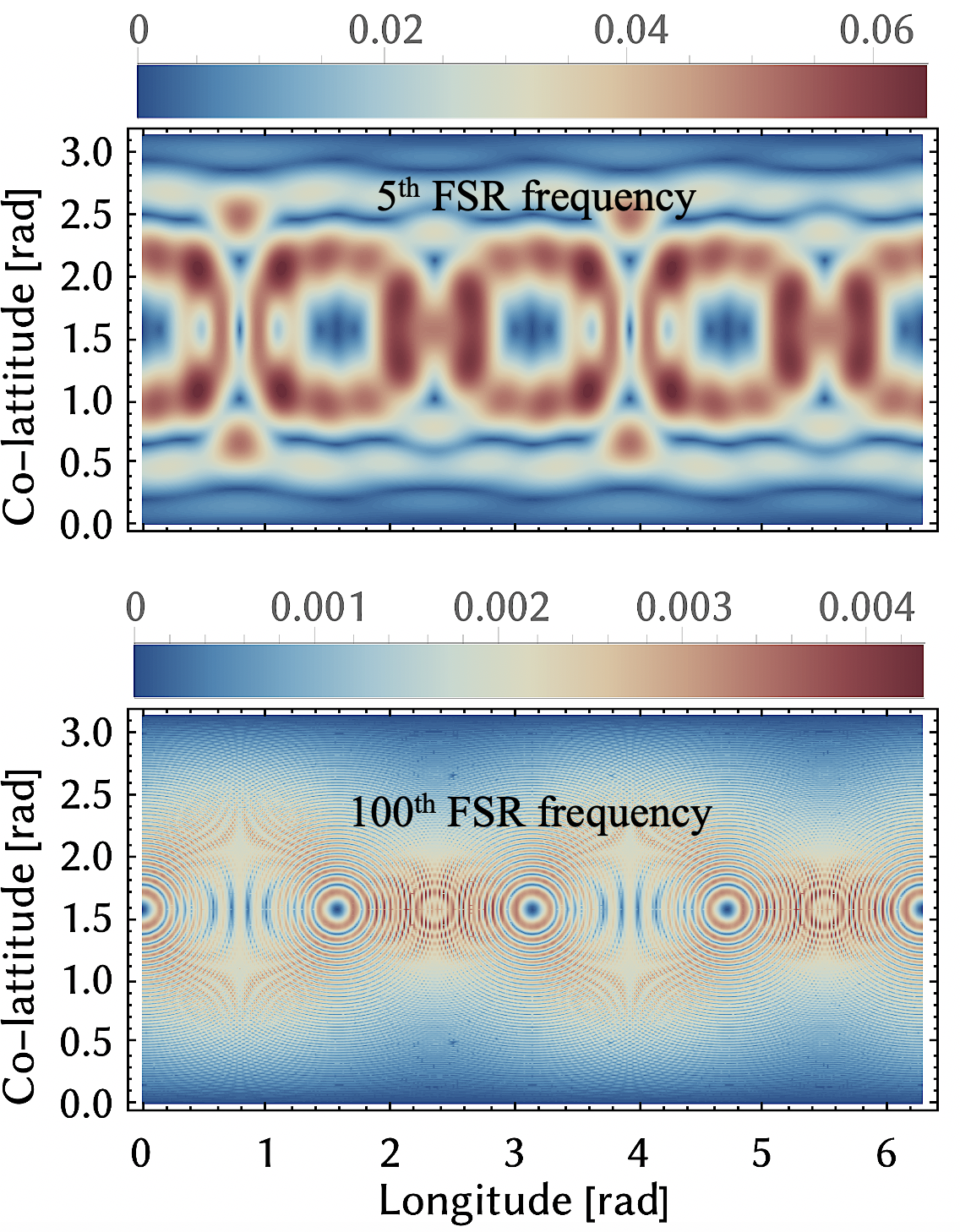}
  \end{minipage}
  \caption{Antennae patterns for detecting gravitational waves of a mixed polarisation at different FSRs: 0th, 1st, 5th and 100th. It can be seen that while for the 0th FSR the maximal response corresponds to the GWs coming from zenith, for higher FSRs the response there is zero.}
  \label{fig:4}
\end{figure}

\section{High-$\textbf{\textit{f}}\!$ antennae pattern}\label{sec:pattern}
Here, we outline how to derive the direction-dependent amplitude of GW induced phase signals for an interferometric GW observatory with perpendicular arms following the work \cite{Rakhmanov2008,Rakhmanov2009}.
The arms are oriented along $x$ and $y$, of which the $x$-arm provides the reference.
We define the GW-induced change of the optical round-trip phase as
\begin{equation} \nonumber
\varphi_a = h_a\mathcal{F}_a(f)G(f) \, ,
\end{equation}
where $h_a$ is the polarisation of the GW coming from direction $a$ defined with respect to the $x$-arm, $\mathcal{F}_a(f)$ is the  response to this GW, and $G(f)$ is the optical transfer function (e.g.~of an arm cavity). Direction $a$ then reads
\begin{align}\nonumber
  & a_x = \sin \theta \sin \phi \, ,\\ \nonumber
  & a_y = \sin \theta \cos \phi \, ,
\end{align}
where $\theta\in [0,\pi]$ is co-latitude, and $\phi\in [0, 2\pi]$ is longitude of the source.
The polarisation along the two arms is related to $(+,\times)$polarisation through the coordinate transformation
\begin{align}
\label{eq:10}
h_{xx}(\theta, \phi) = &h_{+} (\cos^2 \theta \cos^2\!\phi - \sin^2\!\phi)\\ \nonumber
                                 & +2 h_{\times} \cos \theta \sin \phi \cos \phi,\\
h_{yy}(\theta, \phi) = &h_{+} (\cos^2\!\theta \sin^2\!\phi - \cos^2\!\phi)\\ \nonumber
                                 & -2 h_{\times} \cos \theta \sin \phi \cos \phi \, .
\end{align}

Following the detailed derivation of the high-frequency response given in\,\cite{Rakhmanov2008,Rakhmanov2009}, we define the response function 
\begin{multline} \nonumber
  \mathcal{F}_a(f) = \frac{e^{-i\pi f/f_{\rm FSR}}}{2(1 - a^2)\pi f/f_{\rm FSR}} \\ 
  \times\left( \sin(\pi f/f_{\rm FSR}) - a \sin(\pi a f/f_{\rm FSR}) \right. \\ \left. 
  - i a \left( \cos(\pi f/f_{\rm FSR}) - \cos(\pi f a/f_{\rm FSR}) \right) \right) \, .
\end{multline}
For gravitational waves from zenith ($a = 0$) the equation is reduced to Eq.\,(\ref{eq:5}), which is zero for $f=f_{\rm FSR}$. 
For other orientations, however, the above equation is not zero for $f=f_{\rm FSR}$. 

The phase difference between two arms can be expressed in terms of the response function
\begin{equation} \nonumber
  \Delta \varphi = \varphi_x - \varphi_y = h_{xx}\mathcal{F}_x(f)G_x(f) - h_{yy}\mathcal{F}_y(f)G_y(f) \, .
\end{equation}
Typically, the optical response of the arms is identical, $G_x(f)=G_y(f)=G(f)$. Importantly, in detectors with symmetric arms, the response to GWs only depends on the arm lengths, and not on other parameters of optical responses, which is reflected in the fact that $G(f)$ enters as a common factor.

We can re-write the phase difference in terms of the response of the detector to the two polarisations yielding
\begin{equation} \nonumber
  \Delta  \varphi = (h_{+}\mathcal{F}_+(f) - h_{\times}\mathcal{F}_{\times}(f))G(f),
\end{equation}
where $\mathcal{F}_{+}$ can be computed using Eq.\,(\ref{eq:5}) and formally setting $h_{+}=1, h_{\times}=0$ (and vice versa for $\mathcal{F}_{\times}$). We don't explicitly write the resulting bulky expressions in the interest of space. It is common to use the averaged polarisation response defined by
\begin{equation} \nonumber
  \bar{\mathcal{F}} = \sqrt{|\mathcal{F}_{+}|^2 + |\mathcal{F}_{\times}|^2 }.
\end{equation} 
To quantify the sensitivity of a GW observatory, it is also common to use the sky-averaged response. We have followed both in Fig.~\ref{fig:3}.

Alternatively, the signal amplitude can be displayed for a fixed signal frequency as a function of the localisation in the sky, which are the so-called antennae pattern. 
Fig.~\ref{fig:4} show the characteristic antennae pattern of GW observatories with perpendicular equally log arms for the audio band and a selection of their FSR frequencies. The upper plot presents the response to audio band frequencies. It is maximal for GWs coming from zenith and nadir. 
The next plots present the antennae pattern for frequencies $n \!\cdot \! f_{\rm FSR}$, with $n = 1, 5, 100$.
Here, the highest response at FSR frequencies is achieved for other sky locations.
The higher the frequency, the finer the antenna pattern.
The maximum response is inversely proportional to the order $n$ of the FSR.

\section{Conclusions}
Resonator-enhanced laser interferometers with movable test-mass mirrors have significant optical sensitivities to gravitational waves at a comb of higher kHz, MHz and GHz frequencies. State of the art laser interferometer techniques allow for strain normalized sensitivities below $10^{-22}/ \sqrt{\rm Hz}$ around one MHz and $10^{-19}/ \sqrt{\rm Hz}$ around one GHz (Fig.\,\ref{fig:3}) for 100\,m and 1\,m arm lengths, respectively.  
Modifications to the optical systems of existing laser interferometers are not required for use in the high-frequency range. Main modifications would be faster detector electronics and higher sampling rates, which is easy to realise, as well as adapted absolute calibration procedures.
If the one-MHz range was read out by the currently operating Advanced LIGO, a sensitivity of the order $10^{-22}/ \sqrt{\rm Hz}$ would be achieved, which is well inside the range of existing and proposed dedicated instruments, compare table\,1 in \cite{Aggarwal2021}.

However, even these sensitivities are about seven orders of magnitude too low around one MHz and about 15 orders of magnitude too low around one GHz compared to predicted amplitudes of stochastic sources such as cosmic strings or early-universe first-order phase transitions, see Fig.\,1 in \cite{Servant2024}.
Dedicated interferometric detectors as in Fig.\,\ref{fig:3} would require 10 orders of magnitude increased optical power and ten years of data taking to detect these signals at a MHz. At a GHz, even 24 orders of magnitude increased optical power would be required. 

Interestingly, the existence of gravitational waves at MHz and GHz frequencies can even be detected at audio-band frequencies. This is due to the fact that during the passage of transient gravitational wave energy, the local region of space-time slowly accumulates a lasting space-time distortion. This is the nonlinear memory effect of GW bursts \cite{Christodoulou1991,Favata2010}. 

Ref.\,\cite{McNeill2017} analyses the audio-band memory effect of high-frequency gravitational wave bursts and claims that Advanced LIGO would have already measured the memory effect from arbitrarily high gravitational wave frequencies of the order of $6 \cdot 10^{-21} / \sqrt{\rm Hz}$ with a signal-to-noise ratio (S/N) of five if they existed. 
Fig.\,3 in \cite{McNeill2017} claims that a sine-Gaussian burst signal around 100\,MHz that produces a S/N of 5 in a high-frequency detector with a broadband sensitivity of $10^{-22}/ \sqrt{\rm Hz}$ as proposed in \cite{Goryachev2014} would produce an audio-band S/N of greater 1000 in Advanced LIGO. 
On the similar grounds, Ref.\,\cite{Lasky2021} points out that if the observation in \cite{Goryachev2021} at the frequency of 5.5\,MHz came from a GW, the LIGO/Virgo detectors would have registered this signal with an S/N of greater than $10^6$ in the audio band. 

The current concept for measuring gravitational waves in the audio band therefore provides a considerable sensitivity to MHz and GHz gravitational waves, both directly and indirectly. 
On the other hand, the proposed GW amplitudes in these frequency ranges are extremely weak and the potential sources of them are not widely accepted to exist.
Overall, this highlights the necessity of re-evaluating whether it is essential to invest in the development of novel detectors for the high-frequency range.

% ============================================================ ACKNOWLEDGEMENTS
\vspace{15mm}

\textbf{Data availability} ---
The data that support the plots within this paper and other findings of this study are available from the corresponding author upon reasonable request.\\

\textbf{Code availability} ---
The code that supports the findings of this study is available from the corresponding authors upon reasonable request.

~\\

\begin{acknowledgments}
\textbf{Acknowledgments} ---
We thank Rick Savage and Paul Lasky for fruitful discussion.\\
\end{acknowledgments}

\textbf{Author Contributions} ---
R.S. initiated the research, wrote the main manuscript text and prepared Fig. 1. M.K. developed the theoretical description, contributed to the manuscript text and provided the data for the remaining figures. All authors reviewed the manuscript.\\

\textbf{Competing interests} ---
The authors declare no competing interests.\\

\textbf{Additional information} ---
Correspondence and requests for materials should be addressed to R.S.\\

\end{document}